\documentclass[20pt]{article}
\begin{document}

\title{\textbf{Wormholes in Bulk Viscous Cosmology}}
\author{ Mubasher Jamil \\ \\
\small Center for Advanced Mathematics and Physics,
National University of Sciences and Technology\\
\small Peshawar Road, Rawalpindi, 46000, Pakistan\\
\small mjamil@camp.edu.pk}

\maketitle

\begin{abstract}
We investigate the effects of the accretion of phantom energy with
non-zero bulk viscosity onto a Morris-Thorne wormhole. We have found
that if the bulk viscosity is large then the mass of wormhole
increases rapidly as compared to small or zero bulk viscosity.


\end{abstract}
\large
\section{Introduction}
Exotic tunnel like topology in spacetime commonly called
\textit{wormhole} arises as a solution to the Einstein field
equations. A typical two mouth wormhole joins two arbitrary points
either of the same spacetime or two different spacetimes. Morris and
Thorne \cite{moris} suggested the existence of exotic matter for the
stability of a wormhole. The stress energy tensor $T_{\mu\nu}$ of
the exotic matter must violate the Null energy condition
($T_{\mu\nu}u^{\mu}u^{\nu}\geq0$) where $u^{\mu}$ is the future
directed null vector. They concluded that an advanced civilization
can produce a wormhole for interstellar travel by injecting
sufficient amount of exotic matter in it. Recent interest in
wormhole has arose due to the discovery of exotic phantom energy
driving the accelerated expansion of the universe
\cite{Melch,cald1}. It has been proposed that wormhole can be
stabilized by the accretion of phantom energy and it can result in
increasing the size of the wormhole to engulf the observable
universe \cite{diaz1}. We here consider a similar scenario where we
incorporate the effects of bulk viscous stress in our calculations.
The bulk viscosity is quite relevant in physical cosmology as it can
cause expansion of the universe due to its negative pressure
\cite{coli}.  The presence of viscous fluid can also explain the
observed high entropy per baryon ratio in the universe \cite{mis}.
The formulation of the paper is adopted from \cite{diaz1,babi}.

The plan of this paper is as follows. In the second section, we
present the relativistic model of accretion of viscous phantom
energy onto a stationary wormhole. In third section, we consider two
special cases of bulk viscosity in our model. Finally we conclude
our paper.

\section{Accretion onto wormhole}

We consider a stationary and spherically symmetric wormhole
specified by the line element:
\begin{equation}
ds^{2}=-e^{\Phi(r)}+\frac{dr^2}{1-\frac{b(r)}{r}}+r^2(d\theta^2+\sin^2\theta
d\phi^2).
\end{equation}
The above metric has the following properties \cite{raha,viss}: (1)
The redshift function $\Phi(r)$ must be finite for all values of $r$
thus no horizon exists outside the spacetime, (2) the shape function
$b(r)$ must satisfy the following conditions at the throat $r=r_o$:
$b(r_o)=r_o$ and $b^{\prime}(r_o)<1$, (3) $b(r)<r$ for $r>r_o$ and
(4) the spacetime is asymptotically flat i.e.
$\frac{b(r)}{r}\rightarrow0$ as $|r|\rightarrow\infty$.

Outside the wormhole, we assume the spacetime to be
Friedmann-Robertson-Walker containing only one fluid, namely the
phantom energy with non-vanishing bulk viscosity. The fluid is
assumed to fall onto the WH horizon in the radial direction only
which is in conformity with the spherical symmetry of the WH. Thus
the fluid four velocity is $u^{\mu }=(u^{t}(r),u^{r}(r),0,0)$ which
satisfies the normalization condition $u^{\mu }u_{\mu }=1$. The
corresponding stress energy tensor for the exotic phantom energy is
\begin{equation}
T^{\mu \nu }=(\rho +p_{eff})u^{\mu }u^{\nu }+p_{eff}g^{\mu \nu }.
\label{2}
\end{equation}
Here $p_{eff}=p+p_{visc}$, where $p$ is the isotropic pressure and
$p_{visc}=-3\xi H$ is the bulk viscous pressure with $\xi$ is bulk
viscosity. Using the energy momentum conservation
$T^{\mu\nu}_{;\nu}=0$, we get
\begin{equation}
ur^{2}M^{-2}(\rho
+p_{eff})\left(1-\frac{b(r)}{r}\right)^{-1}\sqrt{u^{2}+\frac{b(r)}{r}-1}=C_{1},
\label{3}
\end{equation}
where $u^{r}=u=dr/ds$ is the radial component of the velocity four
vector and $C_{1}$ is a constant of integration. The second constant
of motion is obtained by contracting the four velocity with the
energy momentum conservation equation $u_{\mu }T_{;\nu }^{\mu \nu
}=0$, which gives
\begin{equation}
u^\mu \rho_{,\mu}+(\rho+p_{eff})u_{;\mu}^\mu=0.
\end{equation}
Integration of Eq. (4) gives the second constant of motion
\begin{equation}
ur^{2}M^{-2}\left(1-\frac{b(r)}{r}\right)^{-1}\exp
\left[\int\limits_{\rho _{\infty }}^{\rho _{h}}\frac{d\rho ^{\prime
}}{\rho ^{\prime }+p_{eff}(\rho ^{\prime })}\right]=-A,  \label{4}
\end{equation}
where $A$ is a constant of integration. Also $\rho _{h}$ and $\rho
_{\infty }$ is the energy density of the phantom fluid at the
horizon of the WH, and at infinity respectively. From Eqs. (3) and
(5) we have
\begin{equation}
(\rho
+p_{eff})\left(1-\frac{b(r)}{r}\right)^{-1/2}\sqrt{u^{2}+\frac{b(r)}{r}-1}\exp
\left[-\int\limits_{\rho _{\infty }}^{\rho _{h}}\frac{d\rho ^{\prime
}}{\rho ^{\prime }+p_{eff}(\rho ^{\prime })}\right]=C_{2}, \label{5}
\end{equation}
here $C_{2}=-C_{1}/A=\tilde{A}( \rho _{\infty }+p(\rho _{\infty
}))$. In order to calculate the rate of change of mass of WH
$\dot{M}$, we integrate the flux of the bulk viscous phantom fluid
over the entire WH horizon i.e.
\begin{equation}
\dot{M}=\oint T_t^{r}dS.  \label{6}
\end{equation}
Here $T_t^{r}$ determines the energy momentum flux in the radial
direction only and $dS=\sqrt{-g}d\theta d\varphi $ is the
infinitesimal surface element of the WH horizon. Using Eqs. (3) -
(7), we get
\begin{equation}
\frac{dM}{dt}=-4\pi DM^{2}\sqrt{1-\frac{b(r)}{r}}(\rho +p_{eff}),
\label{7}
\end{equation}
where $D=A\tilde{A}$ is a positive constant. In the asymptotic
regime $r\rightarrow\infty$, we have
\begin{equation}
\dot{M}=-4\pi DM^2(\rho+p_{eff}),
\end{equation}
which clearly demonstrates the increase in mass of the wormhole if
$\rho + p_{eff}<0$ and vanishing in the opposite case.

\section{Accretion of viscous phantom energy}

We now study the evolution of mass of WH in two special cases: (a)
the constant viscosity; and (b) the power law viscosity.

\subsection{Constant bulk viscosity }

For constant viscosity, the evolution of $a(t)$ is given by
\cite{cata}
\begin{equation}
a(t)=a_o\xi_o^{\frac{-2}{3\gamma}}[\xi_o+\gamma H_o
B(t)]^\frac{2}{3\gamma},
\end{equation}
where
\begin{equation}
B(t)\equiv \exp{\left(\frac{3t\xi_o}{2}\right)}-1,
\end{equation}
and $\gamma<0$. The density evolution is given by
\begin{equation}
\rho(t)=\frac{\rho_o\xi_o^2\exp{(3\xi_o t)}}{[\xi_o+\gamma
H_oB(t)]^2}.
\end{equation}
Further, for $\gamma<0$ the big rip singularity occurs in a finite
time at
\begin{equation}
\tau=\frac{2}{3\xi_o}\ln\left(1-\frac{\xi_o}{H_o\gamma}\right).
\end{equation}
Finally, the WH mass evolution is determined by using Eqs. (10)-(12)
in (9) to get
\begin{equation}
M=M_o\left[1-4\pi DM_o\left\{\frac{2\xi_o}{\gamma}
\ln{\left(\frac{\xi_o}{\xi_o+\gamma
H_oB(t)}\right)}+\frac{2H_oB(t)(\xi_o-\gamma H_o)}{\xi_o+\gamma H_o
B(t)} +3\xi_o\ln{\left(\frac{a}{a_o}\right)} \right\}\right]^{-1}.
\end{equation}
We can analyze this expression in some asymptotic limits. Assume
$\xi_o$ is very large i.e. $\xi_o\gg\gamma H_oB(t)$ and
$\xi_o\gg\gamma H_o$ while $a\sim a_o$. Hence Eq. (14) reduces to
\begin{equation}
M\approx M_o[1-8\pi DM_oH_oB(t)]^{-1}.
\end{equation}
Using the approximation $t\sim\xi_o^{-1}$, Eq. (15) becomes
\begin{equation}
M\approx M_o(1-13\pi DM_oH_o)^{-1}.
\end{equation}
which finally becomes
\begin{equation}
M\approx M_o[1+13\pi DM_oH_o+O(H_o^2)].
\end{equation}
Notice that the current value of Hubble parameter is
$H_o\approx2.3\times 10^{-18}s^{-1}$. Due to smallness of $H_o$, its
contribution to higher order terms in Eq. (17) is negligible. One
can see from Eq. (17) that mass of the wormhole increases under the
assumption of large viscosity. Hence the wormhole is perfectly
supported with the viscous phantom energy.

Similarly, Eq. (14) can be analyzed when $\xi_o\ll1$ and
$\xi_o\ll|\gamma H_oB(t)|$, hence we get
\begin{equation}
M\approx M_o\left[ 1-4\pi DM_o\left\{
\frac{2\xi_o}{\gamma}\ln{\left(\frac{\xi_o}{\gamma
H_oB(t)}\right)}+3\xi_o\ln{\left(\frac{a}{a_o}\right)} \right\}
\right]^{-1}
\end{equation}
In the big rip scenario when $a(t)\rightarrow\infty$, the two
quantities in curly brackets in Eq. (18) will be of the same order
of magnitude having opposite sign and hence cancel each other. Thus
Eq. (18) reduces to
\begin{equation}
M\approx M_o.
\end{equation}

\subsection{Power law bulk viscosity }

 Now we consider the bulk viscosity to possess power law dependence upon
density i.e. $\xi=\alpha\rho^s$, where $\alpha$ and $s$ are constant
parameters. Let us take $\xi=\alpha\rho^{1/2}$ as a special case.
Then the scale factor evolves as \cite{cata}
\begin{equation}
a(t)=a_o\left(1-\frac{t}{\tau}\right)^\frac{2}{3(\gamma-\sqrt{3}\alpha)}.
\end{equation}
The density of phantom fluid evolves as
\begin{equation}
\rho(t)=\frac{4}{3\tau^2(\gamma-\sqrt{3}\alpha)^2}\left(1-\frac{t}{\tau}\right)^{-2},
\end{equation}
or in terms of critical density $\rho_{cr}$ as
\begin{equation}
\rho(t)=\rho_{cr}\left(1-\frac{t}{\tau}\right)^{-2}.
\end{equation}
The corresponding big rip time $\tau$ is given by
\begin{equation}
\tau=\frac{2}{3(\sqrt{3}\alpha-\gamma)}H_o^{-1}.
\end{equation}
Finally, the mass evolution of WH is determined by using Eqs.
(20)-(22) in (9), we get
\begin{equation}
M=M_o\left[1-4\pi DM_o\left \{
\gamma\tau\rho_{cr}\left[1-\left(1-\frac{t}{\tau}\right)^{-1}\right]+3\alpha\sqrt{\rho_{cr}}\left[\left(1-\frac{t}{\tau}\right)^{-1}\ln{a}-\ln{a_o}\right]
\right\}\right]^{-1}.
\end{equation}
In the big rip scenario when $t\rightarrow\tau$ and
$a\rightarrow\infty$, the two quantities in the square brackets in
Eq. (24) will be added. This expression gives the growth of the
wormhole if the dominant energy condition $(\rho+p>0)$ is violated.

\section{Conclusion}
Wormholes are tunnel like topological structures supported by the
exotic matter like phantom energy. In our model, we have
incorporated the viscous pressure along the usual isotropic pressure
in the accretion model. Our model predicts the growth of wormholes
by the accretion of bulk viscous phantom energy. For large bulk
viscosity, the increase in the mass of wormhole is large as compared
to small viscosity.

\subsubsection*{\textit{Acknowledgments:}}
I would like to thank Farook Rahaman for useful discussions during
this work.


\end{document}